\documentclass[reqno, 11pt]{amsart}
\usepackage{amssymb}

\usepackage[pdftex]{graphicx}
\usepackage{listings}
\usepackage{multirow}
\usepackage{placeins}
\usepackage{color}
\usepackage{lscape}
\usepackage{dsfont}
\usepackage{bbm}
\usepackage{graphicx}
\usepackage{subcaption}
\usepackage{rotating}
\usepackage{makecell}
\usepackage{rotating}

\usepackage{graphicx}
\usepackage{multirow}
\usepackage{booktabs}
\usepackage{wrapfig, blindtext}
\label{AsianCall}

\usepackage[square,numbers]{natbib}
\setcitestyle{numbers}

\usepackage{lipsum}
\usepackage{booktabs}

\newcommand{\BlackBoxes}{\global\overfullrule5pt}

\BlackBoxes

\newcommand{\R}{\mathbb{R}}

\newcommand{\E}{\mathbb{E}}

\newcommand{\FC}{\mathcal{F}}

\usepackage{bm}

\newcommand{\vb}[1]{\bm{#1}}

\newcommand{\expo}{\operatorname{exp}} 
\newcommand{\loga}{\operatorname{log}} 
\newcommand{\Fc}{\mathcal{F}}

\newcommand{\Q}{{\mathbb Q}}

\newcolumntype{?}{!{\vrule width 1pt}}

\newtheorem{thm}{Theorem}[section]

\newtheorem{lem}[thm]{Lemma}

\newtheorem{cor}[thm]{Corollary}

\theoremstyle{definition}

\numberwithin{equation}{section}

\def\0{\kern0pt\-\nobreak\hskip0pt\relax}

\makeatletter
\AtBeginDocument{%
 \def\@serieslogo{%
 \vbox to\headheight{%
 \parindent\z@ \fontsize{6}{7\p@}\selectfont
 \vss}}}

\def\makeoverbar#1#2#3#4#5#6#7{%
 \setbox0=\hbox{$\m@th#2\mkern#5mu{{}#3{}}\mkern#6mu$}%
 \setbox1=\null \dimen@=#4\fontdimen8#13 \dimen@=3.5\dimen@
 \advance\dimen@ by \ht0 \dimen@=-#7\dimen@ \advance\dimen@ by \wd0
 \ht1=\ht0 \dp1=\dp0 \wd1=\dimen@
 \dimen@=\fontdimen8#13 \fontdimen8#13=#4\fontdimen8#13
 \rlap{\hbox to \wd0{$\m@th\hss#2{\overline{\box1}}\mkern#5mu$}}
 \fontdimen8#13=\dimen@}

\def\mylabel#1#2{{\def\@currentlabel{#2}\label{#1}}}

\makeatother
\title{Pricing of geometric Asian options in the Volterra-Heston model}

\author[F. \smash{Aichinger}]{Florian Aichinger${}^*$}
\address[F. Aichinger]{Institute for Financial Mathematics and Applied Number Theory, Johannes University of Linz, AT-4040 Linz, Austria / Johann Radon Institute for Computational and Applied Mathematics (RICAM), Austrian Academy of Sciences, AT-4040 Linz, Austria}

\email{florianaichinger@aon.at}

\author[S. \smash{Desmettre}]{Sascha Desmettre${}^\dagger$}
\address[S. Desmettre]{Institute for Financial Mathematics and Applied Number Theory, Johannes Kepler University of Linz, AT-4040 Linz, Austria}

\email{sascha.desmettre@jku.at}

\thanks{${}^*$ Institute for Financial Mathematics and Applied Number Theory, University of Linz, AT-4040 Linz, Austria and RICAM, Austrian Academy of Sciences, AT-4040 Linz, Austria}
\thanks{${}^\dagger$ Institute for Financial Mathematics and Applied Number Theory, University of Linz, AT-4040 Linz, Austria}

\begin{document}

\begin{abstract}
    Geometric Asian options are a type of option where the payoff depends on the geometric mean of the underlying asset over a certain period of time.  
    This paper is concerned with the pricing of such options for the class of Volterra-Heston models, covering the rough Heston model.
    We are able to derive semi-closed formulas for the prices of geometric Asian options with fixed and floating strikes for this class of stochastic volatility models. These formulas require the explicit calculation of the conditional joint Fourier transform of the logarithm of the stock price and the logarithm of the geometric mean of the stock price over time.
    Linking our problem to the theory of affine Volterra processes, we find a representation of this Fourier transform as a suitably constructed stochastic exponential, which depends on the solution of a Riccati-Volterra equation.
    Finally, we provide a numerical study for our results in the rough Heston model.
\end{abstract}

\maketitle

\vspace{0.5cm}
\begin{minipage}{14cm}
{\small
\begin{description}
\item[\rm \textsc{ Key words} ]
{\small Volterra-Heston model, Rough volatility; Asian options; Options pricing; Fourier inversion method; affine Volterra processes}

\item[\rm \textsc{ MSC classification (2020)} ]
{\small 45D05, 60B15, 60L20, 91G20}
\end{description}
}
\end{minipage}
\vspace{5mm}

\section{Introduction}\label{sec:Intro}
\vspace{3mm}
The stochastic volatility model of \cite{H93} is nowadays a standard model for pricing financial derivatives.
In contrast to the Black-Scholes model with constant volatility, in the Heston model, the volatility follows itself a stochastic process; in particular, one  obtains the following dynamics for the stock price and its volatility under a risk-neutral measure $\mathbb{Q}$: \vspace{1mm}
\begin{equation}\label{stockprice} 
d S_t = S_t \left( r dt + \sqrt{\nu_t}d B_t^S\right)\,, 
\end{equation}
\begin{equation}\label{eq:Z}
d \nu_t =\kappa (\theta -\nu_t) dt + \sigma \sqrt{\nu_t} d B_t^{\nu}\,, \vspace{1mm}
\end{equation}
where the process $\nu$ with $\nu_0\ge 0$ is a Cox-Ingersoll-Ross process. The constants $\kappa, \theta, \sigma$ are assumed to be positive and satisfy the Feller condition $2\kappa\theta \ge \sigma^2$, $(B^{\nu}_t)$  and $(B^{S}_t)$  are $(\FC_t)$-adapted Brownian motions, where $\Fc_t:=\sigma(B_u^S,B_u^{\nu}:u\leq t)$ such that $\langle B^S,B^{\nu}\rangle_t = \rho t$, and $r$ denotes the risk-free interest rate. This model fixes, on the one hand, the problem of a non-constant volatility smile, which the Black-Scholes model cannot produce, and, on the other hand, is more suitable to reproduce stylized facts of financial data such as the leverage effect and the mean reversion of volatility. Most importantly, the Heston model still gives a (semi-)explicit option pricing formula in terms of the solution of a Riccati ordinary differential equation which is crucial for a (fast) calibration to market data.

In recent times, since the observation was made that the paths of instantaneous volatilities are rougher than established volatility models suggest, cf. \cite{GJR18}, there is a growing research interest in developing new models that better fit empirical data. Therefore, the popular Heston model \cite{H93} was adapted to the rough volatility framework in  \cite{ER16} by using a fractional process with Hurst index $H<\frac{1}{2}$ as driver of the volatility process. Thus, the dynamics of the volatility process \eqref{eq:Z} become
\begin{equation}\label{eq:rough_H}
\nu_t = \nu_0 + \frac{1}{\Gamma(\alpha)} \int_0^t (t-s)^{\alpha-1} \kappa (\theta-\nu_s) ds  + \frac{1}{\Gamma(\alpha)} \int_0^t (t-s)^{\alpha-1} \sigma \sqrt{\nu_s} dB^{\nu}_s\,,
\end{equation}
where $\Gamma$ denotes the Gamma function. The parameter $\alpha\in (\frac{1}{2},1)$ governs the roughness of the paths and is related to the Hurst parameter of the fractional Brownian motion. \cite{ER16} compute the characteristic function of the log-price in the rough Heston models using a link between fractional volatility models and its microstructural foundation, the so-called Hawkes processes. Similar to the classical case, the characteristic function can be calculated (semi-)explicitly in terms of the solution of a fractional Riccati equation.\\
 A more general class of volatility models covering this rough Heston model is obtained by modeling the volatility process as a stochastic Volterra equation of convolution type, \cite{AJ19, KLP18,PJM21}.
 The Volterra-Heston model under a risk-neutral measure $\mathbb{Q}$ is then given as
 \begin{equation}\label{stockprice_V} 
d S_t =  r S_t  dt + \sqrt{\nu_t} S_t d B_t^S\,, 
\end{equation}
 \begin{align}\label{eq:Volterra}
 \nu_t = \nu_0 +\int_0^t \mathcal{K}(t-s)(\kappa(\theta-\nu_s))ds+\int_0^t \mathcal{K}(t-s)\sigma \sqrt{\nu_s}dB_s^{\nu}\,,
 \end{align}
 with integral kernel $\mathcal{K} \in L^{2}([0,T],\mathbb{R})$.
 The constants $\kappa, \theta, \sigma$ are again assumed to be positive, $(B^{\nu}_t)$  and $(B^{S}_t)$ are Brownian motions such that $\langle B^S,B^{\nu}\rangle_t = \rho t$, and $r$ denotes the risk-free interest rate. We consider a filtration $\{\mathcal{F}_t,t\geq 0\}$ where $\Fc_t:=\sigma(B_u^S,B_u^{\nu}:u\leq t)$. Note that if the kernel $\mathcal{K}$ fulfills\vspace{3mm} 
 \begin{equation}
 \begin{aligned}
        &\mathcal{K}\in L_{\operatorname{loc}}^2 (\R_+,\R) \text{ and there is } \gamma\in(0,2] \text{ such that } \int_0^h \mathcal{K}(t)^2 (t) dt = O(h^{\gamma})   \\
        & \text{and } \int_0^T (\mathcal{K}(t+h)-\mathcal{K}(t))^2 dt = O(h^{\gamma}) \text{ for every } T<\infty,
\end{aligned}
\end{equation}
 and the shifted kernels $\Delta_h \mathcal{K}$, where $\Delta_h \mathcal{K}(t):= \mathcal{K}(t+h)$, satisfy the condition\vspace{3mm}
 \begin{equation}\label{condidion_kernel}
    \begin{aligned}
        & \Delta_h\mathcal{K} \text{ is nonnegative, not identically zero, non-increasing and continuous}\\ 
        & \text{on }(0, \infty),\text{ and its resolvent of the first kind } L \text{ is nonnegative and non-}\\ 
        &\text{increasing in that } s \mapsto L([s, s + t])\text{ is non-increasing for all } t \geq 0,
    \end{aligned}
\end{equation}
 then the existence of a unique in law $\R\times \R_+$-valued continuous weak solution $(S,\nu)$ of \eqref{stockprice_V},\eqref{eq:Volterra} is guaranteed by \cite{AJ19}[Theorem 7.1] for any initial condition $(S,\nu)\in \R\times\R_+$. In particular this holds for the fractional integral kernels $\mathcal{K}(t-s) = (t-s)^{\alpha-1}/\Gamma(\alpha)$, $\alpha\in (\frac{1}{2},1)$ in the rough Heston model \eqref{eq:rough_H}. 

Concerning option pricing in general Volterra models only a few explicit results exist. The seminal paper  \cite{BFG16} shows how the rough fractional stochastic volatility  (RFSV) model by  \cite{GJR18} can be used to price claims on both the underlying and integrated variance, however, no explicit call price formula is given yet. This is done in \cite{HL19} for the rough Heston model in the sense of \cite{ER16}, but not for general Volterra kernels. In the context of the rough Heston model, moment explosion was studied in \cite{GGP19}. The next step of complexity is then exotic options, cf. the well-known monograph of  \cite{ZH98}. Using the above-explained forward variance methods along the lines of \cite{KLP18} (cf. \cite{BFFGJR23}), \cite{HJT18} develop efficient Monte Carlo methods and asymptotic approximations for computing option prices and hedge ratios in models where the log-volatility follows a (modulated) Gaussian Volterra processes. Using kernel-based approximation methods, \cite{CPZ22} price American options in the general Volterra Heston model. \cite{FW23} deal with the pricing and hedging of index options in a rough volatility model, using large deviation methods.  To conclude, \cite{TL23} and \cite{SLM23} deal with the numerical and analytical pricing of Basket and Barrier options based on Fourier methods in rough (versions of) Heston models, however once again, not in general Volterra models. The pricing of Asian options, i.e., options where the payoff depends on the mean of the underlying asset over a certain period of time, has yet not been studied for the rough Heston model or the more general Volterra-Heston model.\\
In the classical Heston model, \cite{KW14} derive semi-analytical pricing formulas for Asian options with fixed and floating strikes. First, they establish formulas for the option prices in terms of the conditional joint Fourier transform of the logarithm of the stock price at maturity and the logarithm of the geometric mean of the stock price over a certain time period. 
This approach is based on methods such as numeraire changes and the Fourier inversion formula, which go back to, e.g., \cite{GP51} and \cite{GKR95}. It is important to note that the proofs do not depend on the underlying volatility model. 
The joint Fourier transform is then represented explicitly as an exponential affine function of the volatility. Finding this representation in the classical model involves Itô calculus and heavily relies on the Markovianity of the volatility process \eqref{eq:Z}.\\
This paper presents pricing formulas for fixed- and floating-strike Asian options in the Volterra-Heston model \eqref{stockprice}, \eqref{eq:Volterra}. The pricing formulas in terms of the joint Fourier transform, as derived by \cite{KW14} for the classical Heston model, are still valid in the Volterra-Heston model. 
The challenging part is thus to explicitly calculate this Fourier transform when the volatility of the stock is modeled as a Volterra square-root process given via \eqref{eq:Volterra}. In contrast to the classical case, the process \eqref{eq:Volterra} is not an Itô process anymore; thus, different techniques must be applied in our case.  
Linking our problem to the theory of affine Volterra processes developed in \cite{AJ19}, we show that in the Volterra-Heston model, the joint Fourier transform can be represented as a suitably constructed stochastic exponential in terms of the so-called forward variance process and the solution of a Volterra-Riccati equation. 
As a main result, we then obtain semi-analytic pricing formulas for both fixed- and floating-strike geometric Asian put and call options in the Volterra- Heston model.
These formulas are a generalization of the results derived in \cite{KW14} for the classical Heston model, and we also show that their results can be recovered as a special case of the general Volterra approach, even if the respective representations of the Fourier transform appear to be quite different at first sight. We also provide an extensive numerical study of Asian option prices in the rough Heston model, which is of particular practical interest. Implementing the pricing formulas, we numerically calculate prices for the different types of geometric Asian options for varying strikes, maturities, and roughness levels and compare our results to those of \cite{KW14} in the classical case. We observe that the effect of the roughness on the price is highly dependent on the maturity of the option.

The paper is structured as follows. 
Section \ref{sec:Intro} serves as an introduction to the class of Volterra-Heston models. 
In Section \ref{sec:European Options}, we present a different parametrization of the Volterra-Heston model via a family of Markovian processes and demonstrate how this can be used for pricing European options. In Section \ref{sec: Asian Options}, we introduce geometric Asian options and a general pricing approach linked to the conditional joint Fourier transform of the logarithm of the stock price and the logarithm of the geometric mean of the stock price over time. In Section \ref{sec:joint FT}, we derive an explicit representation of this conditional joint Fourier transform for the Volterra-Heston model. Section \ref{sec: Geom_asian_options} provides semi-closed pricing formulas for fixed and floating strike Asian call and put options in terms of the Fourier transform calculated in the previous section. In Section \ref{sec: equality}, we compare our results to those obtained in the classical Heston model. Finally, Section \ref{sec: numerics} features a numerical study of our results for the rough Heston model. \\

 \vspace{3mm}
\section{European option pricing in the Volterra Heston model}\label{sec:European Options}  
\vspace{3mm}
To start, we give a short overview of how pricing formulas for European options can be obtained in the Volterra-Heston model. Based on Fourier inversion methods, it is possible to obtain semi-closed formulas for European options in terms of the characteristic function of the stock price $S_T$. In the classical Heston model \cite{H93}, determining this characteristic function involves Itô calculus and hence heavily relies on the Markovianity of the processes $S$ and $\nu$ given via \eqref{stockprice} and \eqref{eq:Z}.
The process $\nu$ defined via \eqref{eq:Volterra} does not have to be Markovian, in particular, this is clearly the case for fractional integral kernels in the rough Heston model \eqref{eq:rough_H}. Thus, in order to apply It\^{o} calculus, one considers an equivalent parametrization of the Volterra Heston model in terms of the family of forward variance processes 
\begin{align}\label{eq:forwardV}
    \xi_t(T):=\mathbb{E}^{\mathbb{Q}}(\nu_T|\mathcal{F}_t)\,,
\end{align}
indexed by $T$ (cf. \cite{GKR19}, \cite{KLP18}). It turns out that for fixed $T$, the process $\xi_t(T)$ is again an Itô process. 
As a result of \cite{KLP18}, one can find the dynamics of the forward variance process as follows. 
  \vspace{3mm}
 \begin{lem}[Prop. 3.2 \cite{KLP18}]
 	Let $\xi_t(T)=\mathbb{E}^{\mathbb{Q}}(\nu_T|\mathcal{F}_t)$ and $R_{\kappa}$ be the resolvent of $\kappa \mathcal{K}$, i.e.
 	\[
 	R_{\kappa}\ast(\kappa \mathcal{K})=(\kappa \mathcal{K})\ast R_{\kappa}=\kappa \mathcal{K}-R_{\kappa}.
 	\]
 	Then the dynamics of $\xi_t(T)$ are given by
 	\[
 	d\xi_t(T)=\frac{1}{\kappa}R_{\kappa}(T-t)\sigma\sqrt{\nu_t}dB_t^{\nu}
 	\]
 	with
 	\[
 	\xi_0(T)=\nu_0 (1-\int_0^T R_{\kappa}(s)ds)+\theta \int_0^T R_{\kappa}(s)ds.
 	\]
 \end{lem}
Note that for the fractional kernel $c\frac{t^{\alpha-1}}{\Gamma(\alpha)}$, the resolvent is given by 
\begin{equation}\label{eq_frac_resolvent}
    R(t)=ct^{\alpha-1}E_{\alpha,\alpha}(-ct^{\alpha}),
\end{equation}
where $E_{\alpha,\beta}(z)=\sum\limits_{n=0}^{\infty}\frac{z^n}{\Gamma(\alpha n +\beta)}$ denotes the Mittag-Leffler function.
The forward variance form of the Volterra Heston model in log-price notation under a risk-neutral measure $\mathbb{Q}$  is then given by
 
 \begin{equation}\label{FVHM1}
d\loga S_t=(r-\frac{1}{2}\nu_t) dt + \sqrt{\nu_t}dB_t^{S},
\end{equation}
\begin{equation}\label{FVHM2}
    d\xi_t(T)= \left\{
\begin{array}{ll}
\frac{1}{\kappa}R_{\kappa}(T-t)\sigma\sqrt{\nu_t}dB_t^{\nu} &\,, t<T \\
0 & \,, t\geq T \\
\end{array}
\right.
;\,\xi_0(T)=\nu_0(1-\int_0^T R_{\kappa}(s)ds)+\theta\int_0^T R_{\kappa}(s)ds. 
\vspace{3mm}
\end{equation} 
Here $R_{\kappa}$ is the resolvent of $\kappa \mathcal{K}$, i.e.
    $R_{\kappa}\ast(\kappa \mathcal{K})=(\kappa \mathcal{K})\ast R_{\kappa}=\kappa \mathcal{K}-R_{\kappa}.$
 Note that the variance $\nu_t$ can be recovered from the forward variance process via $\nu_t=\xi_t(t)$.
 The next theorem then ensures that the characteristic function of the log-spot price in the Volterra Heston model can be determined as a suitably constructed exponential in terms of the forward variance and the solution of a Riccati-Volterra equation:
  \vspace{3mm}

 \begin{thm}[Thm. 3.4 \cite{KLP18}]\label{thm: M}
 	Let $\phi$ be the solution of the Riccati-Volterra equation 
 	\[
 	\phi = \mathcal{K}\ast (Q(iu,\phi)-\kappa \phi), \text{ with}
 	\]
 	\[
 	Q(iu,\phi)=-\frac{1}{2}(u^2+iu)+\sigma\rho i u \phi +\frac{\sigma^2}{2}\phi^2.
 	\]
 	Then the auxiliary process $(M_{\tau})_{0\leq \tau \leq T}$ defined as
 	\begin{align}\label{eq:M}
 	M_{\tau}:=\operatorname{exp}\left(iu (\loga S_{\tau}+r(T-\tau))+\int_{\tau}^T\xi_{\tau}(s)Q(iu,\phi(T-s))ds\right)
 	\end{align}
 	is a true martingale and the characteristic function $\psi_{\loga S_T|\Fc_t}$ is given by
 	\[
 	\psi_{\loga S_T|\mathcal{F}_t}(u)=\mathbb{E}^{\mathbb{Q}}(\operatorname{exp}(iu \loga S_T)|\mathcal{F}_t)=\mathbb{E}^{\mathbb{Q}}(M_T|\mathcal{F}_t)=M_t.
 	\]
 \end{thm}
 \vspace{3mm}
\emph{Proof:} The proof follows as a special case of Lemma \ref{lem:aux_M} for $s=0$ and $w=iu$. \qed\vspace{3mm}
Applying the general model-independent call price formula that connects the distribution of the stock price with its Fourier transform (cf., e.g., \cite{BM00}) and using the above explicit representation of the characteristic function yields the following semi-analytic formula for the European call price in the Volterra Heston model. \vspace{3mm}

 \begin{thm}\label{thm:Volterra_Call}
The price $C(0)$ of a European call option with strike price $K$ and maturity $T$ is given by
\begin{equation}\label{eq: Call_char}
C(0)=S_0 \Pi_1- e^{-rT}K\Pi_2, 
\end{equation}
where
\[
\Pi_1=\frac{1}{2}+\frac{1}{\pi}\int_0^{\infty}\operatorname{Re}\left[\frac{e^{-iu\operatorname{ln}(K)}\psi_{\loga S_T}(u-i)}{iu\psi_{\loga S_T}(-i)}\right]du,
\]

\[
\Pi_2=\frac{1}{2}+\frac{1}{\pi}\int_0^{\infty}\operatorname{Re}\left[\frac{e^{-iu\operatorname{ln}(K)}\psi_{\loga S_T}(u)}{iu}\right]du\,,
\]
and $\psi_{\loga S_T}$ is the characteristic function of the logarithmic stock price at time $T$.
 In the general Volterra Heston model~\eqref{eq:Volterra}, the characteristic function $\psi_{\loga S_T}$ is given by
 \begin{equation}\label{eq: char_explicit}
\psi_{\loga S_T}(u)=\operatorname{exp}\left(iu (\loga S_0+rT)+\int_0^TQ(iu,\phi(T-s))\xi_0(s)ds\right),
 \end{equation}
with
\[
\xi_0(s)=\nu_0(1-\int_0^s R_{\kappa}(y)dy)+\theta\int_0^s R_{\kappa}(y)dy. 
\]
 \end{thm}
  \vspace{3mm}

  \emph{Proof:} The proof for the model-independent pricing formula \eqref{eq: Call_char} is standard and relies on a combination of changes of measure and an application of the Fourier inversion formula. For the sake of completeness, a detailed derivation is given in the Appendix. The explicit representation \eqref{eq: char_explicit} of the characteristic function follows immediately from Theorem \ref{thm: M} for the case $t=0$.  
  
\qed

\vspace{3mm}
The result of Theorem~\ref{thm:Volterra_Call} is a generalization of the result presented in \cite{HL19} for the rough Heston model to general kernels $\mathcal{K}$, in case of a constant interest rate. 
In \cite{HL19}, the interest rate can also be a stochastic process.
\vspace{3mm}

\section{Geometric Asian Options}\label{sec: Asian Options}
Asian options are a type of option where the payoff depends on the mean of the underlying asset over a certain period of time. We distinguish between two different types of such options, namely arithmetic and geometric Asian options. Arithmetic Asian options are options where the payoff depends on the arithmetic mean 
\[
A_{[0,T]}=\frac{1}{T}\int_0^T S_u du,
\]
of the underlying over time.
For such options, no explicit closed-form solution is known in the Heston model (cf. \cite{KW14}), which is already the case for the Black-Scholes model. There is a vast literature on the numerical valuation of such arithmetic options in the Heston model (e.g., \cite{V01}, \cite{FH03}, \cite{BK06}), and therefore we focus on finding explicit formulas for Asian options on the geometric mean.  \\
Denote by $G_{[0,T]}$ the geometric mean of the the stock price $S_t$ over the time period $[0,T]$, i.e.
\begin{equation}\label{eq:geom_mean}
G_{[0,T]}=\operatorname{exp}\left(\frac{1}{T}\int_0^T\operatorname{log}S_u du \right).
\end{equation}
There are four types of geometric Asian option contracts: fixed-strike Asian call options, fixed-strike Asian put options, floating-strike Asian call options, and floating-strike Asian put options. 
The payoffs of fixed strike geometric Asian call and put options with strike $K$ and maturity $T$ are given respectively by 
\begin{equation}
\operatorname{max}\{G_{[0,T]}-K,0\},\quad \operatorname{max}\{K-G_{[0,T]},0\}.
\end{equation}
The payoffs of floating strike geometric Asian call and put options with maturity $T$ are given respectively by 
\begin{equation}
\operatorname{max}\{G_{[0,T]}-S_T,0\},\quad \operatorname{max}\{S_T-G_{[0,T]},0\}.\vspace{3mm}
\end{equation}
Starting from a risk-neutral pricing approach, it can be shown (cf. \cite{KW14}) that prices of Asian options can be expressed in terms of the conditional joint Fourier transform $\psi$ of the logarithm of the stock price at maturity and the logarithm of the geometric mean of the stock price over a certain time period
\begin{equation}\label{eq:psi}
\psi_t(s,w) =\E^{\mathbb{Q}}\left[\expo\left(s \loga G_{t,T} + w \loga S_T \right)|\Fc_t\right],
\end{equation}
with
\begin{equation}\label{eq:G_t_T}
    G_{t,T}= \expo\left(\frac{1}{T}\int_t^T \loga S_u du\right). \vspace{3mm}
\end{equation}
Similarly to the derivation of the formula \eqref{eq: Call_char} for European options, this pricing approach is again based on changes of measure and the Fourier inversion formula, which go back to \cite{GP51} and \cite{GKR95} and can be applied independently of the underlying volatility model. 
The main difficulty is thus to find an explicit representation of the Fourier transform \eqref{eq:psi} in the respective market model. In the classical Heston model, \cite{KW14} derive such an explicit representation as an exponential in terms of the solution of a Riccati equation. However, their proof heavily relies on the Markovianity of the classical Heston volatility process.\\
We want to derive pricing formulas for geometric Asian options in the Volterra-Heston model \eqref{stockprice_V}, \eqref{eq:Volterra}. Thus, we must find an explicit representation for \eqref{eq:psi} when the volatility is modeled as a (non-Markovian) Volterra square-root process. This will be done in the next Section. In Section \ref{sec: Geom_asian_options}, we then present semi-closed pricing formulas for fixed- and floating-strike geometric Asian call and put options.

\vspace{3mm}
\section{The joint conditional Fourier transform}\label{sec:joint FT}
\vspace{3mm}
{
In this Section we show that in the Volterra-Heston model \eqref{stockprice_V}, \eqref{eq:Volterra}, the conditional joint Fourier transform \eqref{eq:psi}
can be represented as a suitably constructed exponential in terms of the forward variance and the solution of a Riccati-Volterra equation. We start with the following Lemma, which links our problem to the theory of affine Volterra processes developed in \cite{AJ19}.}
 \vspace{3mm}
\begin{lem}\label{lem:aux_M}
Define a stochastic process $(M_{\tau})_{0\leq \tau \leq T}$ as 
\begin{equation}\label{aux_M}
    M_{\tau}:= \expo(Y_{\tau})
\end{equation}
with
\begin{equation}\label{eq:M_A}
    \begin{aligned}
    & Y_{\tau} = \E^{\mathbb{Q}}\left[\frac{s}{T}\int_0^T \operatorname{log} S_u du + w\operatorname{log}S_T  | \mathcal{F}_{\tau}\right] \\
    & + \frac{1}{2}\int_{\tau}^T \left(\phi_1^2(T-u) + 2\rho\sigma \phi_1(T-u) \phi_2(T-u) +\sigma^2 \phi_{2}^2(T-u)\right) \xi_{\tau}(u)du,
    \end{aligned}
\end{equation} 
 where $\xi_t(s)=\E^{\mathbb{Q}}[\nu_s|\Fc_t]$ is the forward variance process, $\phi_1\in L^2([0,T],\mathbb{C})$ is an affine linear function given by
    \begin{equation}\label{eq:Ricc1}
        \phi_1(t)= s \frac{t}{T}+w,
    \end{equation}
    and $\phi_2\in L^2([0,T],\mathbb{C})$ is the unique solution of the Volterra-Riccati equation
    \begin{equation}\label{eq:Ricc2}
        \phi_2(t)=\int_0^t \mathcal{K}(t-u)\left[\frac{1}{2}\left(\phi_1^2(u)-\phi_1(u)\right)-\kappa \phi_2(u)+\frac{1}{2}\left(\sigma^2\phi_2^2(u)+2\rho\sigma\phi_1(u)\phi_2(u)\right)\right]du
    \end{equation}
on the interval $[0,T]$.   
Then for all $s,w\in \mathcal{D}:=\{(s,w)\in \mathbb{C}^2 : \operatorname{Re}(s)\geq 0, \operatorname{Re}(w)\geq 0, 0\leq \operatorname{Re}(s)+\operatorname{Re}(w)\leq 1\}$, the process $(M_{\tau})_{0\leq \tau \leq T}$ is a true martingale.     
\end{lem}
\vspace{3mm}

\emph{Proof:}
We aim to write the process $(Y_{\tau})_{0\leq\tau\leq T}$ in terms of a two-dimensional affine Volterra process in order to apply results for this class of affine stochastic processes studied in \cite{AJ19}. To this end, we observe that equations \eqref{stockprice_V} and \eqref{eq:Volterra} can be written as

\begin{equation}
\begin{aligned}
    \begin{pmatrix} \loga S \\ \nu \end{pmatrix}_t & = \begin{pmatrix} \loga S \\ \nu \end{pmatrix}_0 + \int_0^t  \begin{pmatrix} 1 & 0 \\ 0 & \mathcal{K}(t-u) \end{pmatrix} \left[ \begin{pmatrix} r \\ \kappa\theta  \end{pmatrix} + \begin{pmatrix} 0 & -\frac{1}{2} \\ 0 & -\kappa \end{pmatrix} \begin{pmatrix} \loga S \\ \nu \end{pmatrix}_u \right]du \\
    & \qquad\qquad \quad\,\,+ \int_0^t  \begin{pmatrix} 1 & 0 \\ 0 & \mathcal{K}(t-u) \end{pmatrix} \sqrt{\nu_u} \begin{pmatrix} \sqrt{1-\rho^2} & \rho \\ 0 & \sigma \end{pmatrix}  \begin{pmatrix} d\tilde{B}_u \\ dB_u^{\nu} \end{pmatrix}\,,
\end{aligned}    
\end{equation}
where $\tilde{B}$ is a Brownian motion independent of $B^{\nu}$.
Defining the matrix valued integral kernel $\vb{\mathcal{K}}$ as $\vb{\mathcal{K}}:=\begin{pmatrix} 1 & 0 \\ 0 & \mathcal{K} \end{pmatrix}$ with scalar kernel $\mathcal{K}$, we see that  $(\loga S,\nu)^{\top}$ solves the Volterra equation
\begin{equation*}
    (\loga S,\nu)_t^{\top} = (\loga S,\nu)_0^{\top} + \int_0^t \vb{\mathcal{K}}(t-u) b((\loga S,\nu)_u^{\top})du + \int_0^t \vb{\mathcal{K}}(t-u) \sigma((\loga S,\nu)_u^{\top}) d(\tilde{B},B^{\nu})_u^{\top},
\end{equation*}
where
\begin{equation*}
    a((\operatorname{log}S,\nu)^{\top}) =\sigma((\operatorname{log}S,\nu)^{\top})\sigma^{\top}((\operatorname{log}S,\nu)^{\top})= A^{0} + \loga S A^1  + \nu A^2 
\end{equation*}
with
\[
A^0= \vb{0}, A^1=\vb{0}, \text{ and } A^2= \begin{pmatrix} 1 & \rho\sigma \\ \rho \sigma & \sigma^2 \end{pmatrix},
\]
as well as
\begin{equation*}
    b((\operatorname{log}S,\nu)^{\top})= b^{0} + \loga S b^1  + \nu b^2 \text{ with } b^0= \begin{pmatrix} r \\ \kappa\theta \end{pmatrix}, b^1= \begin{pmatrix} 0 \\ 0 \end{pmatrix}, b^2= \begin{pmatrix} -\frac{1}{2} \\ -\kappa \end{pmatrix}.
\end{equation*}
Thus $(\operatorname{log}S,\nu)^{\top}$ is an affine process in the sense of \cite{AJ19}.
Equations \eqref{eq:Ricc1} and \eqref{eq:Ricc2} can be written as a two-dimensional Volterra-Riccati equation
\begin{equation}\label{eq: Ricc2D}
    (\phi_1,\phi_2)= (w,0){\vb{\mathcal{K}}} + \left(\left(\frac{s}{T},0\right)+({\phi_1},\phi_2)B+\frac{1}{2}A(({\phi_1},\phi_2))\right)\ast{\vb{\mathcal{K}}}
\end{equation}
where
\begin{equation*}
    B=(b^1,b^2) \text{ and } A((\phi_1,\phi_2))=((\phi_1,\phi_2)A^1(\phi_1,\phi_2)^{\top},(\phi_1,\phi_2)A^2(\phi_1,\phi_2)^{\top})
\end{equation*}
with solution $(\phi_1,\phi_2)\in L^2([0,T],(\mathbb{C}^2)^*)$.
Thus, $Y_{\tau}$ can be expressed in terms of the two-dimensional affine Volterra process $(\loga S,\nu)^{\top}$ and the solution $(\phi_1,\phi_2)$ of \eqref{eq: Ricc2D} as
\begin{equation}
\begin{aligned}
      Y_{\tau} &= \E^{\mathbb{Q}}\left[(\frac{s}{T},0)\ast \begin{pmatrix} \operatorname{log} S \\\nu \end{pmatrix}(T) + (w,0) \begin{pmatrix} \operatorname{log} S \\\nu \end{pmatrix}_T\Big|\mathcal{F}_{\tau}\right]\\
    &\qquad\qquad\qquad\qquad\qquad\qquad\qquad + \frac{1}{2} \int_{\tau}^T (\phi_1,\phi_2)_{T-u}\left[
 \E^{\mathbb{Q}}(\nu_u|\mathcal{F}_{\tau}) \begin{pmatrix}
1 & \rho\sigma \\
\rho\sigma & \sigma^2 
\end{pmatrix}\right] \begin{pmatrix} \phi_1 \\ \phi_2 \end{pmatrix}_{T-u}du.\\
& =  \E^{\mathbb{Q}}\left[(\frac{s}{T},0)\ast \begin{pmatrix} \operatorname{log} S \\\nu \end{pmatrix}(T) + (w,0) \begin{pmatrix} \operatorname{log} S \\\nu \end{pmatrix}_T  \Big|\mathcal{F}_{\tau}\right]\\
&\qquad\qquad\qquad\qquad\qquad\qquad\qquad + \frac{1}{2} \int_{\tau}^T (\phi_1,\phi_2)_{T-u}a(\E^{\mathbb{Q}}[(\loga S,\nu)^{\top}_u|\Fc_{\tau}])(\phi_1,\phi_2)_{T-u}^{\top}du,
\end{aligned}
\end{equation} 
where
\begin{equation*}
    a(\E^{\mathbb{Q}}[(\loga S,\nu)^{\top}_u|\Fc_{\tau}])=A^0 + \E^{\mathbb{Q}}[\loga S_u|\Fc_{\tau}]A^1 + \E^{\mathbb{Q}}[ \nu_u|\Fc_{\tau}]A^2.
\end{equation*}
Thus Theorem 4.3 of \cite{AJ19} for $X=(\loga S,\nu)$, $u=(w,0)$ and $f=(s/T,0)$ can be applied to prove that $M_{\tau}=\expo(Y_{\tau})$ is a local martingale. It remains to show that $M$ is indeed a true martingale. 
We note that for $s,w\in \mathcal{D}$ it holds that for $\phi_1$ defined via \eqref{eq:Ricc1}, we have $\phi_1\in[0,1]$. Since $\operatorname{Re}(u_2)=0$ and $\operatorname{Re}(f_2)=0$, Theorem 7.1 of \cite{AJ19} yields that $\phi_2$ defined via \eqref{eq:Ricc2} has a unique solution on $[0,T]$ which satisfies $\operatorname{Re}(\phi_2)\leq 0$ and $M_{\tau}$ is a true martingale. 

\qed
\vspace{3mm}

The following Lemma ensures the martingale property for a type of processes that appears in the proof of Theorem \ref{thm_FT}. \vspace{3mm}
\begin{lem}\label{lem:P_mart}
    Let $(P_t)_{t\in[0,T]}$ be a stochastic process given by 
    \begin{equation*}
        P_t:=\int_0^t f(u)\sqrt{\nu_u}dB_u,
    \end{equation*}
    where $f\in C^{\infty}([0,T])$ and $B$ is a Brownian motion. Then $P$ is a true martingale on $[0,T]$.
\end{lem}
\vspace{3mm}
\emph{Proof:} $P$ is a local martingale since it is an integral with respect to a Brownian motion. In order to prove that $P$ is a true martingale, we show that $\E^{\mathbb{Q}}(\langle P,P \rangle_t)<\infty$ for all $t\in[0,T]$. This is the case since for every $t\in[0,T]$:\quad
    $\E^{\mathbb{Q}}[\langle P,P \rangle_t] = \E^{\mathbb{Q}}[\int_0^t f^2(u) \nu_u du] \leq C \int_0^t \E^{\mathbb{Q}}[\nu_u] du < \infty$.
\qed
\vspace{3mm}

Given these auxiliary results, we can now proceed to prove that \eqref{eq:psi} has an explicit representation. It will become evident in the next Section that the condition $s,w\in\mathcal{D}$ on the arguments of $\psi$ does not constitute a restriction for deriving the option pricing formulas.
\vspace{3mm}

\begin{thm}\label{thm_FT}
Let  $(s,w)\in \mathcal{D}$. Then for $t\in [0,T]$, the conditional joint Fourier transform $\psi_t$ defined as
\[
\psi_t(s,w) =\E^{\mathbb{Q}}\left[\expo\left(s \loga G_{t,T} + w \loga S_T \right)|\Fc_t\right]
\]
is given by
    \begin{equation}\label{eq:FT}
        \begin{aligned}
            &\psi_t(s,w) 
             = \expo\bigg(s\left(\frac{T-t}{T}\loga S_t + r \frac{(T-t)^2}{2{T}}\right)+ w\left(\loga S_t + r(T-t)\right)\\
            & + \int_t^T \left(\frac{1}{2}\left(\phi_1^2(T-u)-\phi_1(T-u)\right)+\sigma\rho\phi_1(T-u)\phi_2(T-u)+\frac{1}{2}\sigma^2 \phi_2^2(T-u)\right)\xi_t(u)du\bigg),
        \end{aligned}
    \end{equation}
    where $\xi_t(s)$ is the forward variance process, $\phi_1\in L^2([0,T],\mathbb{C})$ is an affine linear function given by \eqref{eq:Ricc1}
    and $\phi_2\in L^2([0,T],\mathbb{C})$ is the solution of the Volterra-Riccati equation \eqref{eq:Ricc2}.
\end{thm}
\vspace{3mm}
\emph{Proof:}
Starting from the definition of $\psi_t$ and $G_{t,T}$ in \eqref{eq:psi} and \eqref{eq:G_t_T}, we have

\begin{equation*}
        \begin{aligned}
            \psi_t(s,w) & := \E^{\mathbb{Q}}\left[\expo\left(s \loga G_{t,T} + w \loga S_T \right)|\Fc_t\right] = \E^{\mathbb{Q}}\left[\expo\left(s \frac{1}{T}\int_t^T \loga S_u du + w \loga S_T \right) |\Fc_t\right]\\
            & = \expo(-s\frac{1}{T}\int_0^t \loga S_u du)\E^{\mathbb{Q}}\left[\expo\left( s \frac{1}{T}\int_0^T \loga S_u du + w \loga S_T \right) |\Fc_t\right].
        \end{aligned}
    \end{equation*}
Since for the process $M$ defined in \eqref{aux_M} it holds that
\begin{equation*}
    M_T= \expo\left(s \frac{1}{T}\int_0^T \loga S_u du + w \loga S_T\right)\,,
\end{equation*}
and $M$ is a martingale by Lemma \ref{lem:aux_M}, we have 
\begin{equation*}
    \E^{\mathbb{Q}}\left[\expo\left( s \frac{1}{T}\int_0^T \loga S_u du + w \loga S_T \right) |\Fc_t\right] = \E^{\mathbb{Q}}[M_T|\Fc_t]=M_t,
\end{equation*}
and thus 

  \begin{equation}\label{eq:psi_calc}
        \begin{aligned}
            \psi_t(w,s) 
            & = \expo(-s\frac{1}{T}\int_0^t \loga S_u du)\expo(\E^{\mathbb{Q}}(w\operatorname{log}S_T +\frac{s}{T}\int_0^T \operatorname{log} S_u du | \mathcal{F}_{t})) \\
            &  \times \expo(\frac{1}{2}\int_{t}^T \left(\phi_1^2(T-u) + 2\rho\sigma \phi_1(T-u) \phi_2(T-u) +\sigma^2 \phi^2(T-u)\right) \xi_{t}(u)du).
        \end{aligned}
    \end{equation}
We simplify the term $\E^{\mathbb{Q}}(\frac{s}{T}\int_0^T \operatorname{log} S_u du + w\operatorname{log}S_T | \mathcal{F}_{t})$. By \eqref{stockprice_V}, we can write $\loga S_T$ as

\begin{equation*}
    \loga S_T = \loga S_t + (T-t)r -\frac{1}{2}\int_t^T \nu_u du + \int_t^T \sqrt{\nu_u}dB_u^S.
\end{equation*}
Thus, for the conditional expectation, we get
\begin{equation*}
\begin{aligned}
    \E^{\mathbb{Q}}(\loga S_T|\Fc_t) &= \loga S_t + (T-t)r - \frac{1}{2}\E^{\mathbb{Q}}(\int_t^T \nu_u du|\Fc_t) + \E(\int_t^T \sqrt{\nu_u}dB_u^S|\Fc_t)\\
    & = \loga S_t + (T-t)r - \frac{1}{2}\int_t^T \xi_t(u) du,
\end{aligned}    
\end{equation*}
where we have used the fact that the process $P_x^1:=\int_0^x \sqrt{\nu_u}dB_u^S$ is a martingale by Lemma \ref{lem:P_mart}.
Similarly, for the integrated log-spot price, we have
\begin{equation*}
\begin{aligned}
    & \int_0^T \loga S_z dz = \int_0^t \loga S_z dz + \int_t^T (\loga S_t + (z-t)r -\frac{1}{2} \int_t^z \nu_u du +\int_t^z \sqrt{\nu_u}dB_u^S)dz \\
    & = \int_0^t \loga S_z dz+(T-t)\loga S_t + \frac{1}{2}(T-t)^2 r - \frac{1}{2}\int_t^T (T-u)\nu_u du + \int_t^T (T-u)\sqrt{\nu_u}dB_u^S,
\end{aligned}    
\end{equation*}
where we have used the stochastic Fubini theorem from \cite{V12}. Note that the stochastic Fubini theorem is applicable since
\[
\int_t^T(\int_t^z |\sqrt{\nu_u}|^2 d\langle B^S,B^S\rangle_u)^{1/2} dz \leq T (\int_0^T \nu_u du)^{1/2}
\]
is finite a.s. Since $P_x^2:=\int_0^x (T-u)\sqrt{\nu_u}dW_u^S$ is again a martingale by Lemma \ref{lem:P_mart}, we have
\begin{equation*}
\begin{aligned}
    \E^{\mathbb{Q}}(\int_0^T\loga S_z dz|\Fc_t) &= \int_0^t \loga S_u du + (T-t) \loga S_t + \frac{1}{2}(T-t)^2 r - \frac{1}{2}\int_t^T (T-u)\xi_t(u) du\,.
\end{aligned}    
\end{equation*}
This gives us
\begin{equation}\label{eq:E_simple}
\begin{aligned}
    & \E^{\mathbb{Q}}(\frac{s}{T}\int_0^T \operatorname{log} S_u du + w\operatorname{log}S_T| \mathcal{F}_t) = \frac{s}{T}\int_0^t \loga S_u du+ s\left(\frac{T-t}{T}\loga S_t + r \frac{(T-t)^2}{2T}\right) \\
    & \qquad\qquad\qquad\qquad\qquad\qquad\qquad + w\left(\loga S_t + r(T-t)\right) - \frac{1}{2}\int_t^T (s\frac{T-u}{T}+w)\xi_t(u) du. 
\end{aligned}    
\end{equation}

Note that $s\frac{T-u}{T}+w = \phi_1(T-u)$. Taking the exponential and inserting \eqref{eq:E_simple} back into \eqref{eq:psi_calc} thus yields

 \begin{equation*}
        \begin{aligned}
            & \psi_t(s,w) = \expo\bigg(s\left(\frac{T-t}{T}\loga S_t + r \frac{(T-t)^2}{2T}\right)+ w\left(\loga S_t + r(T-t)\right)\\
            & + \int_t^T \left(\frac{1}{2}\left(\phi_1^2(T-u)-\phi_1(T-u)\right)+\sigma\rho\phi_1(T-u)\phi_2(T-u)+\frac{1}{2}\sigma^2 \phi_2^2(T-u)\right)\xi_t(u)du\bigg),
        \end{aligned}
    \end{equation*}
which completes the proof 

\qed
\vspace{3mm}
\section{Pricing Geometric Asian Options in the Volterra-Heston Model}\label{sec: Geom_asian_options}
\vspace{3mm}
In this section, we present semi-closed formulas for the prices of geometric Asian call and put options with fixed or floating strikes in the Volterra-Heston model. Combining the general pricing approach that links the price of geometric Asian options to the joint Fourier transform \eqref{eq:psi}
with the explicit representation of this joint Fourier transform \eqref{eq:FT} derived in the previous section, we obtain the following results.

\subsection{Fixed-strike Asian options}

For a fixed-strike geometric Asian call option, i.e., an option with payoff $\operatorname{max}\{G_{[0,T]}-K,0\}$, the price of the option at time $t$ is given as follows.

 \vspace{3mm}
\begin{thm}
The price $C_{[0,T]}(t)$ of a fixed-strike geometric Asian call option with strike $K$ and maturity $T$ and $t\in[0,T]$ is given by
	\vspace{5mm}
	\begin{equation}\label{eq:fixed_call}
    \begin{aligned}
    C_{[0,T]}(t) &= e^{-r(T-t)+\frac{1}{T}\int_0^t \operatorname{log} S_u du}\\
    &\qquad \times \left[\frac{\psi_t(1,0)-K_{t,T}}{2}+\frac{1}{\pi}\int_0^{\infty}\operatorname{Re}\left(\big(\psi_t(1+iz,0)-K_{t,T}\psi_t(iz,0)\big)\frac{e^{-iz \operatorname{log}K_{t,T}}}{iz}\right)dz\right],
 \end{aligned}
	\end{equation}
where
\begin{equation}\label{eq:K_t_T}
K_{t,T}:= K e^{-\frac{1}{T}\int_0^t \operatorname{log} S_udu}
\end{equation}
and the Fourier transform $\psi_t$ is given by \eqref{eq:FT}.
\end{thm}
 \vspace{3mm}
 \emph{Proof:}
The formula \eqref{eq:fixed_call}, which was originally obtained in \cite[Thm. 4.1]{KW14} for the classical Heston model, holds for every model for which the joint Fourier transform \eqref{eq:psi} is known. It remains to prove that we can apply Theorem \ref{thm_FT} to obtain the exponential representation \eqref{eq:FT} of $\psi_t$ in the Volterra-Heston model. This is the case since all arguments $(s,w)$ of $\psi$ that appear in \eqref{eq:fixed_call} fulfill that
$ s,w\in \mathcal{D}=\{(s,w)\in \mathbb{C}^2 : \operatorname{Re}(s)\geq 0, \operatorname{Re}(w)\geq 0, 0\leq \operatorname{Re}(s)+\operatorname{Re}(w)\leq 1\}$.
 \qed
\vspace{3mm}
 
For a fixed-strike geometric Asian put option, i.e., an option with payoff $\operatorname{max}\{K-G_{[0,T]},0\}$, the price of the option at time $t$ is given as follows.
 \vspace{3mm}
\begin{cor}
The price $P_{[0,T]}(t)$ of a fixed-strike geometric Asian put option with strike $K$ and maturity $T$ and $t\in[0,T]$ is given by
	\vspace{5mm}
	\begin{equation}\label{eq:fixed_put}
    \begin{aligned}
    P_{[0,T]}(t) &= e^{-r(T-t)+\frac{1}{T}\int_0^t \operatorname{log} S_u du}\\
    &\qquad \times \left[\frac{K_{t,T}-\psi_t(1,0)}{2}+\frac{1}{\pi}\int_0^{\infty}\operatorname{Re}\left(\big(\psi_t(1+iz,0)-K_{t,T}\psi_t(iz,0)\big)\frac{e^{-iz \operatorname{log}K_{t,T}}}{iz}\right)dz\right],
 \end{aligned}
	\end{equation}
where $K_{t,T}$ is defined in \eqref{eq:K_t_T}
and the Fourier transform $\psi_t$ is given by \eqref{eq:FT}.
\end{cor}
 \vspace{3mm}
\emph{Proof:}
The formula \eqref{eq:fixed_put} follows from \eqref{eq:fixed_call} and the put-call parity (cf. \cite[Cor. 4.2]{KW14}). Theorem \ref{thm_FT} is again applicable since all arguments $(s,w)$ of $\psi$ that appear in \eqref{eq:fixed_put} fulfill $s,w\in\mathcal{D}$ and hence \eqref{eq:FT} holds.
\qed
\vspace{3mm}
 
\subsection{Floating-strike Asian options}

The price of a floating strike geometric Asian call option, i.e., an option with payoff $\operatorname{max}\{G_{[0,T]}-S_T,0\}$, at time $t$, is given as follows.   \vspace{3mm}
\begin{thm}
The price $\tilde{C}_{[0,T]}(t)$ of a floating-strike geometric Asian call option with maturity $T$ and $t\in[0,T]$ is given by
	\vspace{5mm}
	\begin{equation}\label{eq:float_call}
    \begin{aligned}
    \tilde{C}_{[0,T]}(t) &= e^{-r(T-t)} \bigg[\frac{1}{2}\Big(e^{r(T-t)}S_t-e^{(1/T)\int_0^t \loga S_u du}\psi_t(1,0)\Big)\\
    & + \frac{1}{\pi}\int_0^{\infty}\operatorname{Re}\left(\big(e^{(1/T)\int_0^t \loga S_u du}\psi_t(1+iz,-iz)-\psi_t(iz,1-iz)\big)\frac{e^{i(z/T) \int_0^t \loga S_u du}}{iz}\right)dz\bigg],
 \end{aligned}
	\end{equation}
where
$K_{t,T}$ is defined in \eqref{eq:K_t_T}
and the Fourier transform $\psi_t$ is given by \eqref{eq:FT}.
\end{thm}
 \vspace{3mm}
\emph{Proof:}
The formula \eqref{eq:float_call} is derived in \cite[Cor. 4.4]{KW14}). Theorem \ref{thm_FT} is applicable since all arguments $(s,w)$ of $\psi$ that appear in \eqref{eq:float_call} fulfill $s,w\in\mathcal{D}$ and hence \eqref{eq:FT} holds.
\qed
\vspace{3mm}

Again, by the put-call parity, the price
of a floating-strike geometric Asian put option, i.e., an option with payoff $\operatorname{max}\{S_T-G_{[0,T]},0\}$, at time $t$ is given as follows.
 \vspace{3mm}
\begin{cor}
The price $\tilde{P}_{[0,T]}(t)$ of a floating-strike geometric Asian put option with maturity $T$ and $t\in[0,T]$ is given by
	\vspace{5mm}
	\begin{equation}\label{eq:float_put}
    \begin{aligned}
    \tilde{P}_{[0,T]}(t) &= e^{-r(T-t)} \bigg[\frac{1}{2}\Big(e^{(1/T)\int_0^t \loga S_u du}\psi_t(1,0)-e^{r(T-t)}S_t\Big)\\
    & + \frac{1}{\pi}\int_0^{\infty}\operatorname{Re}\left(\big(e^{(1/T)\int_0^t \loga S_u du}\psi_t(1+iz,-iz)-\psi_t(iz,1-iz)\big)\frac{e^{i(z/T) \int_0^t \loga S_u du}}{iz}\right)dz\bigg],
 \end{aligned}
	\end{equation}
where
$K_{t,T}$ is defined in \eqref{eq:K_t_T}
and the Fourier transform $\psi_t$ is given by \eqref{eq:FT}.
\end{cor}
 \vspace{3mm}
 \emph{Proof:}
The formula \eqref{eq:float_put} is derived in \cite[Thm. 4.3]{KW14}). Theorem \ref{thm_FT} is applicable since all arguments $(s,w)$ of $\psi$ that appear in \eqref{eq:float_put} fulfill $s,w\in\mathcal{D}$ and hence \eqref{eq:FT} holds.
\qed
\vspace{3mm}

\section{Consistency with results in the classical Heston model}\label{sec: equality}

The representation \eqref{eq:FT} of the Fourier transform \eqref{eq:psi} in the Volterra Heston model which we derive in Section \ref{sec:joint FT} is of a different structure than the representation derived in the classical Heston model in \cite{KW14}. In this section, we explicitly show that in the case of $\mathcal{K}\equiv 1$, both representations are identical using the calculus of convolutions and resolvents and the theory of Riccati equations. In the following, we will denote the representation \eqref{eq:FT} for the special case $\mathcal{K}\equiv 1$ and $t=0$ as $\psi_0^{VH}(s,w)$ and the classical representation of \cite{KW14} as $\psi_0^{CH}(s,w)$. We start with a detailed overview of the two representations.

\subsection{Volterra approach} For $\mathcal{K}\equiv 1$, the approach developed in Section \ref{sec:joint FT} leads to

  \begin{equation}\label{eq:psi_V}
        \begin{aligned}
            &\psi_0^{VH}(s,w) 
             = \expo\bigg(s\left(\loga S_0 +  \frac{rT}{2}\right)+ w\left(\loga S_0 + rT\right)\bigg)\times\\
            & \expo\bigg( \int_0^T \left(\frac{1}{2}\left(\phi_1^2(T-{\tau})-\phi_1(T-{\tau})\right)+\sigma\rho \phi_1(T-{\tau})\phi_2(T-{\tau})+\frac{1}{2}\sigma^2 \phi_2^2(T-{\tau})\right)\xi_0({\tau})d{\tau}\bigg),
        \end{aligned}
    \end{equation}
where
\[
\xi_0({\tau})=\E^{\mathbb{Q}}[\nu_{\tau}|\Fc_0]=\nu_0(1-\int_0^{\tau} R_{\kappa}(y)dy)+\theta\int_0^{\tau} R_{\kappa}(y)dy
\]   
is the forward variance for the classical Heston volatility ($\mathcal{K}\equiv 1$) at $t=0$, where the resolvent is given by $R_{\kappa}(y)=\kappa e^{-\kappa y}$, $\phi_1\in L^2([0,T],\mathbb{C})$ is an affine linear function given by
\begin{equation*}
        \phi_1({\tau})= s \frac{{\tau}}{T}+w,
\end{equation*}
    and $\phi_2\in L^2([0,T],\mathbb{C})$ is the solution of the Riccati equation
\begin{equation}\label{eq:phi_2}
        \phi_2({\tau})=\int_0^{\tau} \frac{1}{2}\left(\phi_1^2(y)-\phi_1(y)\right)-\kappa \phi_2(y)+\frac{1}{2}\left(\sigma^2\phi_2^2(y)+2\rho\sigma \phi_1(y) \phi_2(y)\right)dy.
\end{equation}

\subsection{Classical approach} The classical approach of \cite{KW14} leads to
\begin{equation}\label{eq: psi_C}
    \psi_0^{CH}(s,w)=\expo(z_0)\times\expo\left(\nu_0 \hat{C}(0) +\hat{D}(0) \right),
\end{equation}
where $\hat{C}$ is the solution of the Riccati equation
\begin{equation}
    \hat{C}'(\tau)= -\hat{B}(\tau) + \kappa \hat{C}(\tau) - \frac{\sigma^2}{2}\hat{C}^2(\tau),\quad \hat{C}(T)=z_4,
\end{equation}
with the function $\hat{B}$ given as 
\begin{equation}
    \hat{B}(\tau)= z_1 (T-\tau)^2 + z_2 (T-\tau) + z_3,
\end{equation}
and $\hat{D}$ is given via
\begin{equation}
    \hat{D}'(\tau)= -\kappa\theta \hat{C}(\tau),\quad \hat{D}(T)=0.
\end{equation}
The coefficients $z_0$, $z_1$, $z_2$ and $z_3$ are given by
\[
z_0=s\left(\loga S_0 + r\frac{T}{2} - \frac{\kappa\theta\rho T}{2\sigma}-\frac{\rho}{\sigma}\nu_0\right) + w \left(\loga S_0 + rT - \frac{\kappa\theta\rho T}{\sigma} -\frac{\rho}{\sigma}\nu_0 \right),
\]
\[
z_1= \frac{s^2 (1-\rho^2)}{2T^2},\, z_2=\frac{s(2\rho\kappa-\sigma)}{2\sigma T}+\frac{sw(1-\rho^2)}{T},\, z_3=\frac{s\rho}{\sigma T}+\frac{w(2\rho\kappa-\sigma)}{2\sigma}+\frac{w^2(1-\rho^2)}{2}, \,z_4=\frac{\rho w}{\sigma}.
\]

\subsection{Equality of the two approaches} We now show that both representations are indeed equal, i.e., $\psi_0^{VH}(s,w)=\psi_0^{CH}(s,w)$. As a first step, we use time-inversion to get a different representation of \eqref{eq: psi_C}.
Defining $C({\tau}):=\hat{C}(T-{\tau})$, we see that $C(0)=\hat{C}(T)=z_4=\frac{\rho}{\sigma}w$ and 
\begin{equation*}
\begin{aligned}
    C'({\tau}) =-\hat{C}(T-{\tau})& =B(T-{\tau})-\kappa \hat{C}(T-{\tau})+\frac{\sigma^2}{2}\hat{C}^2(T-{\tau})\\ &= z_1 {\tau}^2 + z_2 {\tau} + z_3 - \kappa C({\tau}) +\frac{\sigma^2}{2}C^2({\tau}).
\end{aligned}    
\end{equation*}
Thus $C$ fulfills 
\begin{equation}\label{eq: C_Int}
   C({\tau})= \frac{\rho}{\sigma}w + \int_0^{\tau} z_1 y^2 + z_2 y + z_3 - \kappa C(y) + \frac{\sigma^2}{2}C^2(y) dy.
\end{equation}
Similarly, defining $D({\tau})=\hat{D}(T-{\tau})$, we obtain $D(0)=\hat{D}(T)=0$ and
\[
D'({\tau})=-\hat{D}'(T-{\tau})=\kappa\theta \hat{C}(T-{\tau})=\kappa\theta C({\tau})
\]
and hence $D$ is given by
\[
D({\tau})=\int_0^{\tau} \kappa\theta C(y)dy.
\]
Therefore we obtain with $C$ given via \eqref{eq: C_Int}:
\begin{equation}
    \psi_0^{CH}(s,w)=\expo(z_0)\times\expo\left(\nu_0 C(T) +\kappa\theta\int_0^T C({\tau}) d{\tau} \right)\,.
\end{equation}
As a next step, we simplify the convolution appearing in \eqref{eq:psi_V} to show that 
\[
\psi_0^{VH}(s,w) = \expo\bigg(s\left(\loga S_0 + r \frac{T}{2}\right)+ w\left(\loga S_0 + rT\right)\bigg)\times \expo\left(\nu_0 \phi_2(T)+\kappa\theta \int_0^T \phi_2({\tau})d{\tau}\right).
\]
To this end, we define
\begin{equation}\label{eq:QQ}
Q({\tau}):=\frac{1}{2}\left(\phi_1^2({\tau})-\phi_1({\tau})\right)+\sigma\rho \phi_1({\tau})\phi_2({\tau})+\frac{1}{2}\sigma^2 \phi_2^2({\tau})\,.
\end{equation}
Thus, we can write
\[
\psi_0^{VH}(s,w) = \expo\bigg(s\left(\loga S_0 + r \frac{T}{2}\right)+ w\left(\loga S_0 + rT\right)\bigg)\times \expo\left(\int_0^T Q(T-{\tau})\xi_0({\tau})d{\tau}\right).
\]
The result now follows since 
\begingroup
\allowdisplaybreaks
\begin{align*}
        \int_0^T Q(T-{\tau})\xi_0({\tau})d{\tau} &=\int_0^T Q(T-{\tau})\left(\nu_0 - \nu_0 \int_0^{\tau} R_{\kappa}(z)dz +\theta \int_0^{\tau} R_{\kappa}(z)dz\right)d{\tau}\\
        & = \nu_0 \int_0^T Q(T-{\tau}) d{\tau} - \nu_0 Q\ast \vb{1} \ast R_{\kappa}(T) + \theta Q\ast \vb{1} \ast R_{\kappa}(T)\\
        & = \nu_0 \int_0^T Q({\tau}) d{\tau} - \nu_0 \vb{1}\ast R_{\kappa}\ast Q (T) + \theta \vb{1}\ast  R_{\kappa}\ast Q  (T)\\
        & = \nu_0\left(\phi_2(T)+\kappa \int_0^T \phi_2({\tau})d{\tau}\right) - \nu_0 \kappa \int_0^T \phi_2({\tau})d{\tau} +\kappa\theta \int_0^T \phi_2({\tau})d{\tau}\\
        & = \nu_0 \phi_2(T) + \kappa\theta \int_0^T \phi_2({\tau}) d{\tau}.
\end{align*}
\endgroup    
Here, we have used the commutativity of convolution for the third equality. For the fourth equality, we used the identity $\phi_2=\frac{1}{\kappa}R_{\kappa}\ast Q$ as well as the fact that $\phi_2(T)=\int_0^{T} Q(\tau) -\kappa \phi_2(\tau)d\tau$ which follows by \eqref{eq:phi_2} and definition \eqref{eq:QQ}. A short proof of this identity is provided in the Appendix. In the next step, we show that the solutions of the Riccati equations \eqref{eq: C_Int} and \eqref{eq:phi_2} are related to each other. More precisely, we show that if $C({\tau})$ is the solution of \eqref{eq: C_Int}, then $\phi_2({\tau}):= C({\tau})-\frac{\rho}{\sigma}\phi_1({\tau})$ is the solution of \eqref{eq:phi_2}. This is true since
\begingroup
\allowdisplaybreaks
\begin{align*}
        & C({\tau})-\frac{\rho}{\sigma}\phi_1({\tau}) = \int_0^{\tau} \frac{1}{2}(\phi_1^2(y)-\phi_1(y))-\kappa(C(y)-\frac{\rho}{\sigma}\phi_1(y))\\
        & \qquad\qquad\qquad\qquad\qquad\qquad\qquad\qquad +\rho\sigma \phi_1(y)(C(y)-\frac{\rho}{\sigma}\phi_1(y))+\frac{\sigma^2}{2}(C(y)-\frac{\rho}{\sigma}\phi_1(y))^2 dy\\
        & \iff C({\tau})-\frac{\rho}{\sigma}\phi_1({\tau}) = \int_0^{\tau} \frac{1}{2}(\phi_1^2(y)-\phi_1(y))+\frac{\rho\kappa}{\sigma}\phi_1(y)-\frac{\rho^2}{2}\phi_1^2(y) - \kappa C(y)+\frac{\sigma^2}{2}C^2(y) dy\\
        & \iff C({\tau})-\frac{\rho}{\sigma}\phi_1({\tau}) = \int_0^{\tau} z_1 y^2 + z_2 y +z_3 - \frac{\rho s}{\sigma T} - \kappa C(y)+\frac{\sigma^2}{2}C^2(y) dy\\
        & \iff C({\tau})-\frac{\rho}{\sigma}(\frac{s{\tau}}{T}+w) = - \frac{\rho}{\sigma}\frac{s{\tau}}{T} + \int_0^{\tau} z_1 y^2 + z_2 y +z_3 - \kappa C(y)+\frac{\sigma^2}{2}C^2(y) dy\\
        & \iff C({\tau}) =  \frac{\rho}{\sigma}w + \int_0^{\tau} z_1 y^2 + z_2 y +z_3 - \kappa C(y)+\frac{\sigma^2}{2}C^2(y) dy.
\end{align*}
\endgroup    
Finally, we combine the results of the two previous steps to obtain
\vspace{0.2cm}
\begingroup
\allowdisplaybreaks
\begin{align*}
            \psi_0^{VH}(s,w) 
             & = e^{s\left(\loga S_0 + r \frac{T}{2}\right)+ w\left(\loga S_0 + rT\right)} \cdot e^{\nu_0 \phi_2(T)+\kappa\theta \int_0^T \phi_2({\tau})d{\tau}}\\
             & = e^{s\left(\loga S_0 + r \frac{T}{2}\right)+ w\left(\loga S_0 + rT\right)}  \cdot e^{\nu_0 (C(T)-\frac{\rho}{\sigma}(s+w))+\kappa\theta \int_0^T C({\tau})-\frac{\rho}{\sigma}(\frac{s}{T}{\tau}+w) d{\tau}}\\
             & = e^{s\left(\loga S_0 + r \frac{T}{2}\right)+ w\left(\loga S_0 + rT\right)} \cdot e^{-s\frac{\rho}{\sigma}\nu_0 - w\frac{\rho}{\sigma}\nu_0 - s\frac{\kappa\theta\rho T}{2\sigma}-w\frac{\kappa\theta\rho T}{\sigma}}\cdot e^{\nu_0 C(T)+\kappa\theta \int_0^T C({\tau}) d{\tau}}\\
             & = e^{s\left(\loga S_0 + r\frac{T}{2} - \frac{\kappa\theta\rho T}{2\sigma}-\frac{\rho}{\sigma}\nu_0\right) + w \left(\loga S_0 + rT - \frac{\kappa\theta\rho T}{\sigma} -\frac{\rho}{\sigma}\nu_0 \right)}
              \cdot e^{\nu_0 C(T)+\kappa\theta \int_0^T C({\tau}) d{\tau}}\\
             & = e^{z_0}\cdot e^{\nu_0 C(T) +\kappa\theta\int_0^T C({\tau}) d{\tau}} =  \psi_0^{CH}(s,w).
 \end{align*}
\endgroup    
\vspace{0.2cm}
This is the desired equality. \qed

\section{Numerical results}\label{sec: numerics}

In this section, we present numerical results for fixed- and floating-strike geometric Asian call and put options for the class of fractional Heston models, i.e. we consider the model \eqref{stockprice_V}, \eqref{eq:Volterra} for the fractional integral kernel $K(t)=\frac{t^{\alpha-1}}{\Gamma(\alpha)}$. For $\alpha=1$, we recover the classical Heston model, and hence, in this special case, our results correspond to those derived in \cite{KW14}. Note that the parameter $\alpha$, which governs the roughness of the paths, is related to the Hurst parameter $H$ of the fractional Brownian motion via $H=\alpha-\frac{1}{2}$. 
As already lined out, since the observation was made that volatility is rough (cf. \cite{GJR18}), the class of rough Heston models with $\alpha\in (\frac{1}{2},1)$, has become of particular theoretical and practical interest, and thus we will put particular emphasis on the impact of the roughness parameter on the option price. 

\subsection{Useful representation of the option prices}

To start off, we summarize the pricing formulas derived in the previous Sections for the case $t=0$. 
\vspace{3mm}

\begin{cor}
    For $t=0$, the price of a fixed-strike geometric Asian call option is given by
    \begin{equation}\label{eq: C_0 fixed}
    \begin{aligned}
    C_{[0,T]}(0) &= e^{-rT} \left[\frac{\psi_0(1,0)-K}{2}+\frac{1}{\pi}\int_0^{\infty}\operatorname{Re}\left(\big(\psi_0(1+iz,0)-K\psi_0(iz,0)\big)\frac{e^{-iz \operatorname{log}K}}{iz}\right)dz\right],
 \end{aligned}
	\end{equation}
 the price of a fixed-strike geometric Asian put option is given by
 \begin{equation}\label{eq: P_0 fixed}
    \begin{aligned}
    P_{[0,T]}(0) &= e^{-rT} \left[\frac{K-\psi_0(1,0)}{2}+\frac{1}{\pi}\int_0^{\infty}\operatorname{Re}\left(\big(\psi_0(1+iz,0)-K\psi_0(iz,0)\big)\frac{e^{-iz \operatorname{log}K}}{iz}\right)dz\right],
 \end{aligned}
	\end{equation}
 the price of a floating-strike geometric Asian call option is given by
\begin{equation}\label{eq: C_0 float}
    \begin{aligned}
    \tilde{C}_{[0,T]}(0) &= e^{-rT} \bigg[\frac{e^{rT}S_0-\psi_0(1,0)}{2} + \frac{1}{\pi}\int_0^{\infty}\operatorname{Re}\left(\big(\psi_0(1+iz,-iz)-\psi_0(iz,1-iz)\big)\frac{1}{iz}\right)dz\bigg],
 \end{aligned}
	\end{equation}
 the price of a floating-strike geometric Asian put option is given by
\begin{equation}\label{eq: P_0 float}
    \begin{aligned}
    \tilde{P}_{[0,T]}(0) &= e^{-rT} \bigg[\frac{\psi_0(1,0)-e^{rT}S_0}{2} + \frac{1}{\pi}\int_0^{\infty}\operatorname{Re}\left(\big(\psi_0(1+iz,-iz)-\psi_0(iz,1-iz)\big)\frac{1}{iz}\right)dz\bigg],
 \end{aligned}
	\end{equation}
and the conditional joint Fourier transform $\psi_0$ can be written as
 \begin{equation}\label{eq: psi_0}
        \begin{aligned}
            &\psi_0(s,w) 
             = \expo\bigg(s\left(\loga S_0 +  \frac{rT}{2}\right)+ w\left(\loga S_0 + rT\right)+ \int_0^T \nu_0Q({\tau}) + \kappa(\theta -\nu_0) \phi_2({\tau}) d{\tau} \bigg),
        \end{aligned}
    \end{equation} 
    where $\phi_2\in L^2([0,T],\mathbb{C})$ is the solution of the Volterra Riccati equation
\begin{equation}\label{eq: Ricc}
        \phi_2({\tau})=\int_0^{\tau} K({\tau}-y)\left(\frac{1}{2}\left(\phi_1^2(y)-\phi_1(y)\right)-\kappa \phi_2(y)+\frac{1}{2}\left(\sigma^2\phi_2^2(y)+2\rho\sigma \phi_1(y) \phi_2(y)\right)\right)dy,
\end{equation}    
$\phi_1\in L^2([0,T],\mathbb{C})$ is an affine linear function given by
\begin{equation*}
        \phi_1({\tau})= s \frac{{\tau}}{T}+w,
\end{equation*}
and $Q$ is defined as
\[
Q({\tau})=Q(\phi_1,\phi_2;{\tau}):=\frac{1}{2}\left(\phi_1^2({\tau})-\phi_1({\tau})\right)+\frac{1}{2}\left(\sigma^2\phi_2^2({\tau})+2\rho\sigma \phi_1({\tau}) \phi_2({\tau})\right). \vspace{3mm}
\]
\end{cor}

\emph{Proof:} The formulas \eqref{eq: C_0 fixed}, \eqref{eq: P_0 fixed}, \eqref{eq: C_0 float} and \eqref{eq: P_0 float} are immediately obtained by setting $t=0$ in equations \eqref{eq:fixed_call}, \eqref{eq:fixed_put} \eqref{eq:float_call},  and \eqref{eq:float_put}. 
For the representation of the Fourier transform $\psi_0$, we first observe that the formula \eqref{eq:FT} evaluated at $t=0$ yields
\[
\psi_0(s,w) = \expo\bigg(s\left(\loga S_0 + r \frac{T}{2}\right)+ w\left(\loga S_0 + rT\right)\bigg)\times \expo\left(\int_0^T Q(T-{\tau})\xi_0({\tau})d{\tau}\right).
\]
Substituting  $\xi_0({\tau})=\nu_0(1-\int_0^{\tau} R_{\kappa}(y)dy)+\theta\int_0^{\tau} R_{\kappa}(y)dy$ and applying the calculus of convolutions and resolvents to the last term yields
\vspace{3mm}
\begin{equation*}
    \begin{aligned}
        \int_0^T Q(T-{\tau})\xi_0({\tau})d{\tau} &=\int_0^T Q(T-{\tau})\left(\nu_0 - \nu_0 \int_0^{\tau} R_{\kappa}(z)dz +\theta \int_0^{\tau} R_{\kappa}(z)dz\right)d{\tau}\\
        & = \nu_0 \int_0^T Q(T-{\tau}) d{\tau} - \nu_0 Q\ast \vb{1} \ast R_{\kappa}(T) + \theta Q\ast \vb{1} \ast R_{\kappa}(T)\\
        & = \nu_0 \int_0^T Q({\tau}) d{\tau} - \nu_0 \vb{1} \ast R_{\kappa}\ast Q (T) + \theta \vb{1} \ast  R_{\kappa}\ast Q  (T)\\
        & = \nu_0 \int_0^T Q({\tau}) d{\tau} - \nu_0 \kappa \int_0^T \phi_2({\tau})d{\tau} +\kappa\theta \int_0^T \phi_2({\tau}){\tau}.\vspace{3mm}
    \end{aligned}
\end{equation*}

Here, we have used the commutativity of convolution for the third and the identity $\phi_2=\frac{1}{\kappa}R_{\kappa}\ast Q$ (cf. Lemma \ref{lem: Ricc_equ} in the Appendix) for the fourth equality. This now gives us \eqref{eq: psi_0}. 
\qed   
\vspace{3mm}

Note that in contrast to  \eqref{eq:FT}, the resolvent does no longer appear in the representation  \eqref{eq: psi_0} of $\psi_0$ anymore. This fact is particularly useful in the case of fractional integral kernels appearing in the rough Heston model since we do not have to approximate the Mittag-Leffler function appearing in the resolvent (cf. equation \eqref{eq_frac_resolvent}). Therefore, this allows us to improve the accuracy as well as the computational efficiency of the implementation of our formulas. \vspace{3mm}

\subsection{Concrete results} 

We now present concrete numerical results for fixed- and floating-strike geometric Asian calls and put option prices at $0$. In order to have a benchmark parameter set, analogous to \cite{KW14}, we use the calibrated market parameters
\[\kappa=1.15\,,\, \theta=0.348 \,,\, \sigma=0.39 \mbox{ and } \rho=-0.64\,,\]  which correspond to the daily averages of the respective implied parameters reported in table IV of \cite{BCC97} as well as their choices
\[r=0.05\,,\, S_0=100 \mbox{ and } \nu_0=0.09\,.\]
Two approximations are carried out for the numerical implementation of the analytic formulas.
In order to calculate the Fourier transform \eqref{eq: psi_0}, one first has to numerically solve the Volterra Riccati equation \eqref{eq: Ricc} for $K(t)=\frac{t^{\alpha-1}}{\Gamma(\alpha)}$, which is done using the fractional Adams method, developed in \cite{DFF02}. For a detailed error analysis of the fractional Adams method, we refer to \cite{DFF04}.
Secondly, the infinite integral appearing in \eqref{eq: C_0 fixed}, \eqref{eq: C_0 float} must be truncated. Here, we take an upper limit of $10^2$. This is in line with the observation made in \cite{KW14}, that for higher limits the numerical results remain unchanged. 

In Table \ref{tab:fixed_call}, we present prices for fixed-strike geometric Asian call options for different maturities, strike prices, and roughness levels of volatility. In our study, we investigate the case $\alpha\in (\frac{1}{2},1)$, which is the rough Heston model. 

The case, where $\alpha\in (1,\frac{3}{2})$, leading to smoother volatility paths than in the classical model is not considered, since in this case the kernel is not completely monotone. In particular condition \eqref{condidion_kernel} does not hold and the existence, uniqueness, and affine property is not guaranteed.

We observe the following characteristics: The option price decreases with increasing strike. Moreover, the smaller the maturity, the higher the impact of the strike on the option price. Larger maturities lead to higher option prices across all different strikes. All these observations align with the results of \cite{KW14} for the classical Heston model. Note in particular that for $\alpha=1$ in Table \ref{tab:fixed_call}, we recover their results presented in \cite[Table 5]{KW14}. 

In addition to the classical Heston model, we now face a further parameter that has to be taken into account, namely the roughness parameter $\alpha$. We observe that the effect of the roughness depends on the maturity of the option. For smaller maturities, the price of the option increases with the roughness level. 

For the fixed-strike put option prices in Table \ref{tab:fixed_call}, the option price rises with increasing strike and maturity. As is also the case for the corresponding call options, only for very large maturities do prices start to decline slightly.  
The effect of roughness again changes depending on the maturity of the option. Here, the price of the option again increases with the roughness for smaller maturities. However, we observe the reverse effect for larger maturities, i.e., option prices decrease with increasing roughness levels. 
This is in line with the fact that time-dependent effects of the roughness are well-established in the context of portfolio optimization theory, where the preference between buying a rough or a smooth stock changes with the time-horizon (cf. \cite{PJM21}, \cite{GH20}, \cite{AD23}). However, in addition, the prices listed in the corresponding tables are related to each other via the put-call parity
\[
C_{[0,T]}(0)-P_{[0,T]}(0) = e^{-rT}(\psi_0(1,0)-K).
\]
The value of the Fourier transform $\psi_0$ also depends on the roughness parameter $\alpha$, and hence, the effects of roughness on put and call option prices cannot be expected to be observed on a one-for-one basis. Please note that the put prices have, at the same time, been checked for correctness with the help of this put-call parity.

The results for floating-strike geometric Asian call and put options given in Table \ref{tab:float_call} exhibit a similar behavior. Larger maturities again lead to higher option prices for both put and call options. As for fixed-strike options, the effect of the roughness depends on the maturity of the option. For our model parameters, we observe that a change in the effect of roughness appears around $T=1$. For maturities $T<1$, rougher volatility paths lead to higher option prices, while the reverse effect can be observed for $T>1$ for both put and call options. Note that in contrast to the fixed-strike case, the stock price at maturity $S_T$, which appears in the payoff, is dependent on the roughness parameter as well.
Here, the prices listed in the corresponding tables are related to each other via the put-call parity
\[
\tilde{C}_{[0,T]}(0)-\tilde{P}_{[0,T]}(0) = S_0 - e^{-rT}\psi_0(1,0)\,,
\]
and have once more been checked for correctness using this parity.

\begin{table}
    \centering
   \resizebox{12.9cm}{!}{\begin{tabular}{|c|r||r|r|r||r|r|r|} \hline 
 \multicolumn{2}{|c||}{}&\multicolumn{3}{|c||}{Call option value}&\multicolumn{3}{|c|}{Put option value}\\ \hline 
         T (years)&  K&  $\alpha=1.00$&  $\alpha=0.75$& $\alpha=0.60$ & $\alpha=1.00$&  $\alpha=0.75$& $\alpha=0.60$\\ \hline
 0.2&  90&  10.6571&  10.8151& 10.9546 & 0.4526&  0.6424& 0.8061\\ \hline 
 &  95&    6.5871&  6.7931& 6.9556 & 1.3329&  1.5705& 1.7575\\ \hline  
 & 100&  3.4478& 3.6182&3.7484 & 3.1438& 3.3459& 3.5005\\ \hline 
 &  105&    1.4552&  1.5338& 1.5976 & 6.1014&  6.2118& 6.3000\\ \hline  
 &  110&   0.4724&  0.4841& 0.5016& 10.0689&  10.1124& 10.1542\\ \hline  \hline
 
 0.4&  90&   11.7112&  11.9773& 12.1707& 1.3799&  1.7123& 1.9469\\ \hline  
 &  95&    8.0894&  8.3894& 8.5983& 2.6591&  3.0254& 3.2755\\ \hline  
 &  100&  5.1616&  5.4415& 5.6345& 4.6323&  4.9785& 5.2128\\ \hline  
 &  105&    3.0018&  3.2178& 3.3695& 7.3735&  7.6558& 7.8489\\ \hline  
 &  110&    1.5715&  1.7089& 1.8103& 10.8442&  11.0478& 11.1906\\ \hline  \hline
 
 0.5& 90& 12.2329& 12.5314&12.7380&  1.8619& 2.2389& 2.4909\\ \hline 
 & 95&  8.7553& 9.0844&9.3050&  3.2609& 3.6685& 3.9345\\ \hline 
 & 100&  5.8971& 6.2119&6.4211&  5.2792& 5.6726& 5.9273\\ \hline 
 & 105&  3.7072& 3.9697&4.1460&  7.9659& 8.3069& 8.5288\\ \hline 
 & 110&  2.1589& 2.3501&2.4823&  11.2941& 11.5638& 11.7417\\ \hline \hline
 
 1& 90&  14.5779& 14.9390&15.1588&  4.1801& 4.6239& 4.8798\\ \hline 
 & 95&  11.5551& 11.9353&12.1637&  5.9135& 6.3763& 6.6408\\ \hline 
 & 100&  8.9457& 9.3239&9.5499& 8.0602& 8.5211& 8.7830\\ \hline 
 & 105&  6.7559& 7.1130&7.3263&  10.6266& 11.0663& 11.3155\\ \hline 
 & 110&  4.9723& 5.2932&5.4859& 13.5991& 14.0027& 14.2312\\ \hline \hline
 
 1.5& 90&  16.5030& 16.8533&17.0544&  6.2606& 6.6283& 6.8267\\ \hline 
 & 95&  13.7625& 14.1213&14.3261& 8.1588& 8.5350& 8.7371\\ \hline 
 & 100&  11.3374& 11.6939&11.8969& 10.3725& 10.7464& 10.9465\\ \hline 
 & 105&  9.2245& 9.5687&9.7647& 12.8983& 13.2599& 13.4529\\ \hline 
 & 110&  7.4122& 7.7358&7.9205&  15.7246& 16.0658& 16.2473\\ \hline \hline
 
 2& 90&  18.0914& 18.4083&18.5845&  8.0965& 8.3323& 8.4548\\ \hline 
 & 95&  15.5640& 15.8803&16.0558&  10.0933& 10.3285& 10.4502\\ \hline 
 & 100&  13.2933& 13.6034&13.7755&  12.3467& 12.5758& 12.6939\\ \hline 
 & 105&  11.2728& 11.5719&11.7379& 14.8504& 15.0684& 15.1804\\ \hline 
 & 110&  9.4921& 9.7761&9.9339&  17.5939& 17.7968& 17.9005\\ \hline \hline
 
 3& 90&  20.5102& 20.7529&20.8857&  11.1123& 11.0617& 11.0298\\ \hline 
 & 95&  18.3060& 18.5346&18.6598&  13.2116& 13.1469& 13.1073\\ \hline 
 & 100&  16.2895& 16.5034&16.6208& 15.4986& 15.4193& 15.3717\\ \hline 
 & 105&  14.4531& 14.6523&14.7617&  17.9658& 17.8716& 17.8159\\ \hline 
 & 110&  12.7882& 12.9725&13.0740&  20.6044& 20.4954& 20.4316\\ \hline\hline

  8& 90&  24.7980& 24.9009& 24.9967&  18.0809& 17.3813& 16.9256\\ \hline 
 & 95&  23.4852& 23.5573& 23.6319&  20.1198& 19.3894& 18.9125\\ \hline 
 & 100&  22.2483& 22.2920& 22.3471&  22.2345& 21.4757& 20.9792\\ \hline 
 & 105&  21.0827& 21.1003& 21.1374&  24.4204& 23.6356& 23.1211\\ \hline 
 & 110&  19.9840& 19.9778& 19.9983& 26.6733& 25.8646& 25.3336\\ \hline\hline

12& 90&  24.1840& 24.3052& 24.4514&  18.9998& 18.2823& 17.7561\\ \hline 
 & 95&  23.2584& 23.3520& 23.4766&  20.8183& 20.0731& 19.5254\\ \hline 
 & 100&  22.3795& 22.4472& 22.5516&  22.6834& 21.9124& 21.3445\\ \hline 
 & 105&  21.5442& 21.5878& 21.6734&  24.5922& 23.7971& 23.2103\\ \hline 
 & 110&  20.7498& 20.7710& 20.8388&  26.5418& 25.7243& 25.1198\\ \hline

    \end{tabular}}
    \caption{Fixed-strike geometric Asian call and put option prices for different maturities, strike prices, and roughness levels for model parameters $S_0=100$, $\nu_0=0.09$, $t=0$, $r=0.05$, $\kappa=1.15$, $\theta=0.348$, $\sigma=0.39$.}
    \label{tab:fixed_call}
\end{table}
\vspace{3mm}

\begin{table}
    \centering
   \resizebox{12.9cm}{!}{\begin{tabular}{|c||r|r|r||r|r|r|} \hline 
 \multicolumn{1}{|c||}{}&\multicolumn{3}{|c||}{Call option value}&\multicolumn{3}{|c|}{Put option value}\\ \hline 
         T (years)&    $\alpha=1.00$&  $\alpha=0.75$& $\alpha=0.60$ & $\alpha=1.00$&  $\alpha=0.75$& $\alpha=0.60$\\ \hline
 0.2&    3.9925&  4.2467& 4.3812&   3.3015&  3.5240& 3.6341\\ \hline  
 0.4&    6.4473&  6.7508& 6.8805&    4.9965&  5.2336& 5.3222\\ \hline  
 0.5&  7.5727& 7.8563& 7.9658&   5.7216&  5.9267& 5.9908\\ \hline 
 1&   12.5761&  12.5965& 12.5621&   8.5845&  8.5222& 8.4516\\ \hline  
 1.5&   16.8305&  16.5500& 16.3745&   10.5698&  10.2720& 10.0988\\ \hline  
 2&    20.5478&  20.0250& 19.7323&    11.9781&  11.5364& 11.2964\\ \hline  
 3&    26.8612&  26.0414& 25.5759&    13.7228&  13.1961& 12.8935 \\ \hline  
 8&    48.0585&  47.0888& 46.3269&    15.1044&  14.9363& 14.7224\\ \hline  
 12&    59.2926&  58.4512& 57.6810 &   13.8699&  13.8658& 13.7619\\ \hline  

    \end{tabular}}
    \caption{Floating-strike geometric Asian call and put option prices for different maturities and roughness levels for model parameters $S_0=100$, $\nu_0=0.09$, $t=0$, $r=0.05$, $\kappa=1.15$, $\theta=0.348$, $\sigma=0.39$.}
    \label{tab:float_call}
\end{table}
\vspace{3mm}

\section{Conclusion}
We provide semi-analytic pricing formulas for fixed- and floating-strike geometric Asian options in the class of Volterra-Heston models, which include the rough Heston model. These formulas are obtained by combining a general pricing approach based on the Fourier inversion theorem with an explicit representation of the conditional joint Fourier transform of the logarithm of the stock price at maturity and the logarithm of the geometric mean of the stock price over a certain time period. In contrast to the classical Heston model, the volatility process in the Volterra-Heston model is no longer Markovian, and hence, we have to apply different techniques to obtain this representation as a suitably constructed stochastic exponential in terms of the solution of a Volterra-Riccati equation. This construction requires a link to the theory of affine Volterra processes developed in \cite{AJ19}. Our work is an extension of the results of \cite{KW14} for the classical Heston model to the class of Volterra-Heston models. We also implement our pricing formulas and provide a numerical study for the rough Heston model, in which we investigate the influence of the roughness of the volatility process on the option prices.

\section*{Acknowledgements}
F. Aichinger is supported by the Austrian Science Fund (FWF) Project P34808 (Grant-DOI: 10.55776/P34808). For the purpose of open access, the authors have applied a CC BY public copyright license to any author-accepted manuscript version arising from this submission. 

\section{Appendix}

\subsection{Proof of the general pricing approach in Theorem \ref{thm:Volterra_Call}:}Under a risk-neutral measure $\Q$, the price of the European call option is given as
\begin{equation*}
    \begin{aligned}
        C_0 & = e^{-rT}\mathbb{E}^{\mathbb{Q}}(\operatorname{max}(S_T-K,0))\\
        & = e^{-rT}[\E^{\Q}(S_T \mathbbm{1}_{\{S_T>K\}})-\E^{\Q}(K \mathbbm{1}_{\{S_T>K\}})].
    \end{aligned}
\end{equation*}
\vspace{3mm} 
For the second expectation, we get 
\[
\E^{\Q}(K \mathbbm{1}_{\{S_T>K\}})= K \Q(S_T>K).
\]
Define a probability measure $\Q_S$ via $\Q_S(A):=\E^{\Q}\left(\frac{S_T}{\E^{\Q}(S_T)}\mathbbm{1}_A\right)$. Then, for the first expectation we obtain

\begin{equation*}
    \begin{aligned}
        \E^{\Q}(S_T \mathbbm{1}_{\{S_T>K\}}) & = \E^{\Q}\left(\E^{\Q}(S_T)\frac{S_T}{\E^{\Q}(S_T)} \mathbbm{1}_{\{S_T>K\}}\right)\\ 
        & = \E^{\Q}(S_T) \E^{\Q}\left(\frac{S_T}{\E^{\Q}(S_T)} \mathbbm{1}_{\{S_T>K\}}\right) = e^{rT}S_0 \Q_S(S_T>K).
    \end{aligned}
\end{equation*}
Thus we have
\begin{equation}\label{eq:C_0}
    \begin{aligned}
        C_0 & = S_0 \Q_S(S_T>K) - e^{-rT} K \Q(S_T>K)\\
         & = S_0 \Q_S(\loga S_T>\loga K) - e^{-rT} K \Q(\loga S_T> \loga K)\\
    \end{aligned}
\end{equation}
Using the identity
\begin{equation*}
    \begin{aligned}
        \psi_{\loga S_T}^{\Q_S}(u) & = \E^{\Q_S}(e^{iu \loga S_T}) = \E^{\Q}\left(e^{iu \loga S_T} \frac{d \Q_S}{d\Q}\right) \\
        & = \E^{\Q}\left(e^{iu \loga S_T} \frac{S_T}{\E^{\Q}(S_T)}\right) = \frac{ \E^{\Q}(e^{iu \loga S_T} S_T)}{{\E^{\Q}(S_T)}} = \frac{\psi_{\loga S_T}^{\Q}(u-i)}{\psi_{\loga S_T}^{\Q}(-i)},
    \end{aligned}
\end{equation*}
an application of the Fourier inversion formula now yields
\begin{equation}\label{eq: Pi_1}
    \begin{aligned}
        \Q_S(\loga S_T> \loga K) & = \frac{1}{2}+ \frac{1}{\pi} \int_0^{\infty} \operatorname{Re}\left[\frac{e^{-iu \loga K}\psi_{\loga S_T}^{\Q_S}(u)}{iu}\right] du \\
         & = \frac{1}{2}+\frac{1}{\pi}\int_0^{\infty}\operatorname{Re}\left[\frac{e^{-iu\operatorname{log}(K)}\psi_{\loga S_T}^{\Q}(u-i)}{iu\psi_{\loga S_T}^{\Q}(-i)}\right]du,
    \end{aligned}
\end{equation}
and
\begin{equation}\label{eq: Pi_2}
\Q(\loga S_T>\loga K)= \frac{1}{2}+ \frac{1}{\pi} \int_0^{\infty} \operatorname{Re}\left[\frac{e^{-iu \loga K}\psi_{\loga S_T}^{\Q}(u)}{iu}\right] du.
\end{equation}
Inserting \eqref{eq: Pi_1} and \eqref{eq: Pi_2} into equation \eqref{eq:C_0} yields the desired result

\qed

\subsection{Equivalent form of the Riccati-Volterra equation}
We derive an equivalent representation for the Riccati-Volterra equations appearing in Sections \ref{sec: equality} and \ref{sec: numerics}.

\begin{lem}\label{lem: Ricc_equ}
The Riccati-Volterra equation
\[
\phi = K\ast (Q(\phi,f)-\kappa \phi)
\]
with
\[
Q(\phi,f)=\frac{1}{2}\left(f^2-f\right)+\frac{1}{2}\left(\sigma^2\phi^2+2\rho\sigma f \phi\right)\vspace{3mm}
\]
has equivalent representation 
\[
\phi = \frac{1}{\kappa}R_{\kappa}\ast Q(\phi,f).
\]
\end{lem}

\emph{Proof:}
We start with the Volterra-Riccati equation
\[
\phi = K\ast (Q(\phi,f)-\kappa \phi)
\]
and subtract $R_{\kappa}\ast \phi$ from both sides. This yields

\begingroup
\allowdisplaybreaks
\begin{align*}
            \phi - R_{\kappa}\ast \phi 
             & =  K\ast (Q(\phi,f)-\kappa \phi) - R_{\kappa}\ast K \ast (Q(\phi,f)-\kappa \phi)\\
             & =  (K - R_{\kappa}\ast K) \ast  (Q(\phi,f)-\kappa \phi) \\
             & = \frac{1}{\kappa} R_{\kappa} \ast (Q(\phi,f)-\kappa \phi) = \frac{1}{\kappa}R_{\kappa}\ast Q(\phi,f) - R_{\kappa}\ast \phi.
 \end{align*}
\endgroup    

 And consequently, we get the equivalent representation 
 \[
 \phi = \frac{1}{\kappa}R_{\kappa}\ast Q(\phi,f).
 \]

 \qed

\end{document}